\documentclass[aps,prd,floatfix,nofootinbib,twocolumn]{revtex4}

\usepackage{graphicx}
\usepackage{dcolumn}
\usepackage{bm}
\usepackage{amssymb}
\usepackage{amsmath}
\usepackage[caption = false]{subfig}
\usepackage{float}
\usepackage{hyperref}

\newcommand{\fdm}{f_\mathrm{DM}}
\newcommand{\Mx}{M_\chi}
\newcommand{\sigmaxp}{\sigma_{\chi\mathrm{p}}}
\newcommand{\phihat}{\hat{\phi}}
\newcommand{\MESAstar}{\texttt{MESA star}}
\newcommand{\MESA}{\texttt{MESA}}
\newcommand{\gev}{\mathrm{GeV}}
\newcommand{\mev}{\mathrm{MeV}}
\newcommand{\kev}{\mathrm{keV}}
\newcommand{\cm}{\mathrm{cm}}
\newcommand{\tcool}{t_\mathrm{cool}}
\newcommand{\sigmaCrit}{\sigmaxp^\mathrm{crit}}
\newcommand{\Rsun}{R_\mathrm{sun}}
\newcommand{\kpc}{\mathrm{kpc}}
\newcommand{\msolar}{\mathrm{M}_\odot}
\newcommand{\sigman}{\sigma_N}
\newcommand{\vesc}{v_\mathrm{esc}}
\newcommand{\sigmav}{\langle\sigma_{\mathrm{a}}\mathrm{v}\rangle}

\begin{document}

\title{Indirect Probes of Dark Matter and Globular Cluster Properties From Dark Matter Annihilation within the Coolest White Dwarfs} 
\author{Travis J. Hurst, Andrew R. Zentner, Aravind Natarajan, and Carles Badenes}
\email{TJH47@PITT.edu}
\affiliation{Department of Physics and Astronomy, \& The Pittsburgh Particle physics, Astrophysics, and Cosmology Center (Pitt-PACC), University of Pittsburgh, Pittsburgh, PA 15260, USA}
\date{\today}

\begin{abstract}

White Dwarfs (WD) capture Dark Matter (DM) as they orbit within their host halos. These captured particles may subsequently annihilate, heating the stellar core and preventing the WD from cooling. The potential wells of WDs are considerably deeper and core temperatures significantly cooler than those of main sequence stars. Consequently, DM evaporation is less important in WDs and DM with masses $M_{\chi} \gtrsim 100\, \kev$ and annihilation cross-sections orders of magnitude below the canonical thermal cross-section ($\sigmav \gtrsim 10^{-46}\, \cm^3$/s) can significantly alter WD cooling in particular astrophysical environments. We consider WDs in globular clusters (GCs) and dwarf galaxies. If the parameters of the DM particle are known, then the temperature of the coolest WD in a GC can be used to constrain the DM density of the cluster's halo (potentially even ruling out the presence of a halo if the inferred density is of order the ambient Galactic density). Recently several direct detection experiments have seen signals whose origins might be due to low mass DM. In this paper, we show that if these claims from CRESST, DAMA, CDMS-Si, and CoGeNT could be interpreted as DM, then observations of NGC 6397 limit the fraction of DM in that cluster to be $f_{\mathrm{DM}} \lesssim 10^{-3}$. This would be an improvement over existing constraints of 3 orders of magnitude and clearly rule out a significant DM halo for this cluster. More generally, we show how such observations constrain combinations of DM and GC properties. Building on previous work, we also show how observations of WDs in dwarf galaxies, such as Segue I, can provide novel constraints on low mass DM or DM with very low contemporary annihilation cross-sections as may be realized in models in which s-wave annihilation is suppressed and p-wave annihilation dominates. This paper provides further motivation for high-quality observations of stellar populations as a probe of dark matter.
 
\end{abstract}
\maketitle

\section{Introduction}
\label{section:intro}

In the current cosmological paradigm, $\approx 23\%$ of the energy of the Universe is in the form of Dark Matter (DM) \cite{WMAP9,Planck}. The nature of this DM is unknown, but the most popular particle candidates are Weakly-Interacting Massive Particles (WIMPs). Among other reasons, WIMPs are compelling candidates because they can be produced thermally in the early Universe such that their relic abundance yields the correct value for the DM density ($\Omega_{\mathrm{DM}} \approx 0.23$). Moreover, WIMPs arise naturally in various extensions to the Standard Model (SM) of particle physics. As such, detecting, identifying, or excluding broadly WIMP-like DM is a high priority in cosmology and particle physics. In this paper, we describe the implications of recent signals in a variety of dark matter search experiments for the structure of globular cluster NGC 6397 and its DM content.
%
%
 We also follow up on previous work \cite{Bertone}, by demonstrating how observations of cool, dim WDs, in concert with future measurements of dark matter properties, can place unique and powerful constraints on the DM content of globular clusters.

It is now well known that a star will capture weakly-interacting DM particles through scattering off of nuclei within the stellar interior as the star orbits within the host DM halo (see e.g. the review~\cite{Jungman} and references therein). DM particles captured in this way will accumulate within the star and achieve sufficiently high densities such that annihilation of the DM particles can come to equilibrium with capture (see \S\ref{section:annihilation}). The annihilations will be a source of energy for the stellar interior. Consequently, there should be some heating from DM annihilation in the cores of stars \cite{Burners,Burners2}. The effects of such annihilations have been considered for main sequence (MS) stars \cite{Burners3}, primordial stars \cite{Spolyar, Natarajan, Iocco}, brown dwarfs \cite{zentner_hearin11}, and compact stars \cite{Hooper,Bertone}. 
%
%
If WDs are heated by DM annihilations in their cores, they will be prevented from cooling below some minimum temperature set by the equilibration of energy injection from annihilations with energy loss by radiation \cite{Hooper}. Observations of cool, dim WDs lead to constraints on DM because the energy injected by DM annihilation cannot exceed the observed WD luminosity.

There are a number of well-developed efforts to identify the dark matter particle. Direct detection experiments seek to identify DM by observing scattering events in a detector on Earth. Several direct detection experiments have reported intriguing results recently including the Coherent Germanium Neutrino Technology (CoGeNT)~\cite{CoGent}, the silicon detectors of the Cryogenic Dark Matter Search (CDMS-Si)~\cite{CDMS-Si}, Cryogenic Rare Event Search with Superconducting Thermometers~(CRESST) \cite{CRESST} and Dark Matter Large sodium Iodide Bulk for RAre processes (DAMA/LIBRA, henceforth just DAMA)~\cite{DAMA}. These experiments all see events that are inconsistent with expected backgrounds. Taking all of these experiments together, there is no consistent interpretation of these events as DM particle scattering (though there are some models that can alleviate the tension between the experiments, e.g. Refs.~\cite{Kelso,Foot,Zurek,Panci1,Panci2,Arina}).

Broadly speaking these anomalous signals point toward a DM particle with a mass of $\Mx \sim 5$-$25\, \gev$ and a cross-section for scattering off of a nucleon of $\sigmaxp \sim 10^{-41}\, \cm^2$, though the details of the viable mass ranges and cross-sections depend upon the particular experiment under consideration (the relevant areas of parameter space are shown in Fig.~\ref{fig:mvsigma} below). Recent results from the Large Underground Xenon Dark Matter (LUX) \cite{LUX} and SuperCDMS \cite{SuperCDMS} experiments exclude nearly all of the parameter space consistent with a DM interpretation of the anomalous events in the aforementioned experiments, casting serious doubt on the viability of interpreting these events as due to DM. As an example, Ref.~\cite{Zurek} found that even considering non-standard models of WIMP DM scattering, such as an Anapole interaction, which can bring the regions of interest for CDMS-Si, CoGeNT, and DAMA into alignment, the parameter space preferred by these experiments is excluded by the LUX limits.

A second avenue of inquiry into the nature of DM is indirect detection. Indirect detection experiments seek to identify the DM by observing the particles (e.g. photons, neutrinos, positrons) produced by the self-annihilation of DM in specific astrophysical environments, such as the Sun \cite{Modak} (for which the relevant annihilation product is the neutrino) or the Galactic Halo (for which the relevant annihilation products may be photons \cite{Fermi}, neutrinos \cite{IceCube}, or positrons \cite{AMS}). Several groups have found evidence that there is an anomalous, extended gamma-ray emission coming from the Galactic Center using data from the Large Area Telescope aboard the Fermi Gamma Ray Space Telescope (Fermi-LAT). This emission is consistent with the annihilation of a relatively low-mass ($\Mx \sim$ 5-30~GeV) DM particle to quarks \cite{Abazajian, Good&Hooper, Hooper&Linden, Abaz&Kap, Gordon&Mac, Mac&Gordon} or leptons \cite{Abazajian, Hooper&Good, Hooper&Linden, Abaz&Kap, Gordon&Mac, Mac&Gordon, Kyae}, but the evidence is not conclusive (e.g. \cite{Boyarsky}).  Fermi-LAT has also observed the dwarf spheroidal satellite galaxies of the Milky Way.  The lack of a significant $\gamma$-ray signal from these DM dominated objects places constraints on annihilating DM in the mass range 2 Gev - 10 TeV \cite{Fermi2,Geringer, Cholis}. 
%
%
As an example, new analysis places an upper-limit on the annihilation rate which is below the canonical thermal value for $\Mx$ = 6-35 GeV depending on the annihilation channel \cite{Geringer2}.

The anomalous signals mentioned above suggest that it may not be long until some of the basic properties of the DM particle are determined. Identification of DM scattering cross-sections opens the door to what some authors have referred to as {\em WIMP astronomy} (e.g., Ref.~\cite{Peter}). The premise of WIMP astronomy is that with the properties of DM known, observations of astronomical phenomena, such as cosmic ray fluxes or stellar evolution, can yield further information about the structure and substructure of the Milky Way's Galactic DM halo. Improved knowledge of the structure of the Galactic DM halo can, in turn, lead to improved constraints on the processes that lead to the formation of the Galactic halo and the Milky Way Galaxy or Milky Way sub-systems (e.g., dwarf galaxies \cite{Hooper}, globular clusters \cite{Bertone}). 
%
%
Here we follow up on Ref.~\cite{Bertone} by showing the implications of these recent anomalous signals, interpreted as DM, for the DM content of the Milky Way globular cluster NGC 6397.  
%
%
Because the signals coming from indirect detection experiments may not be due to DM, we also show the reach of WD cooling as a tool for WIMP astronomy as a function of the properties of the DM particle in a general context.

%
\begin{figure}[htp]
\centering
\includegraphics[width=9cm, height=9cm]{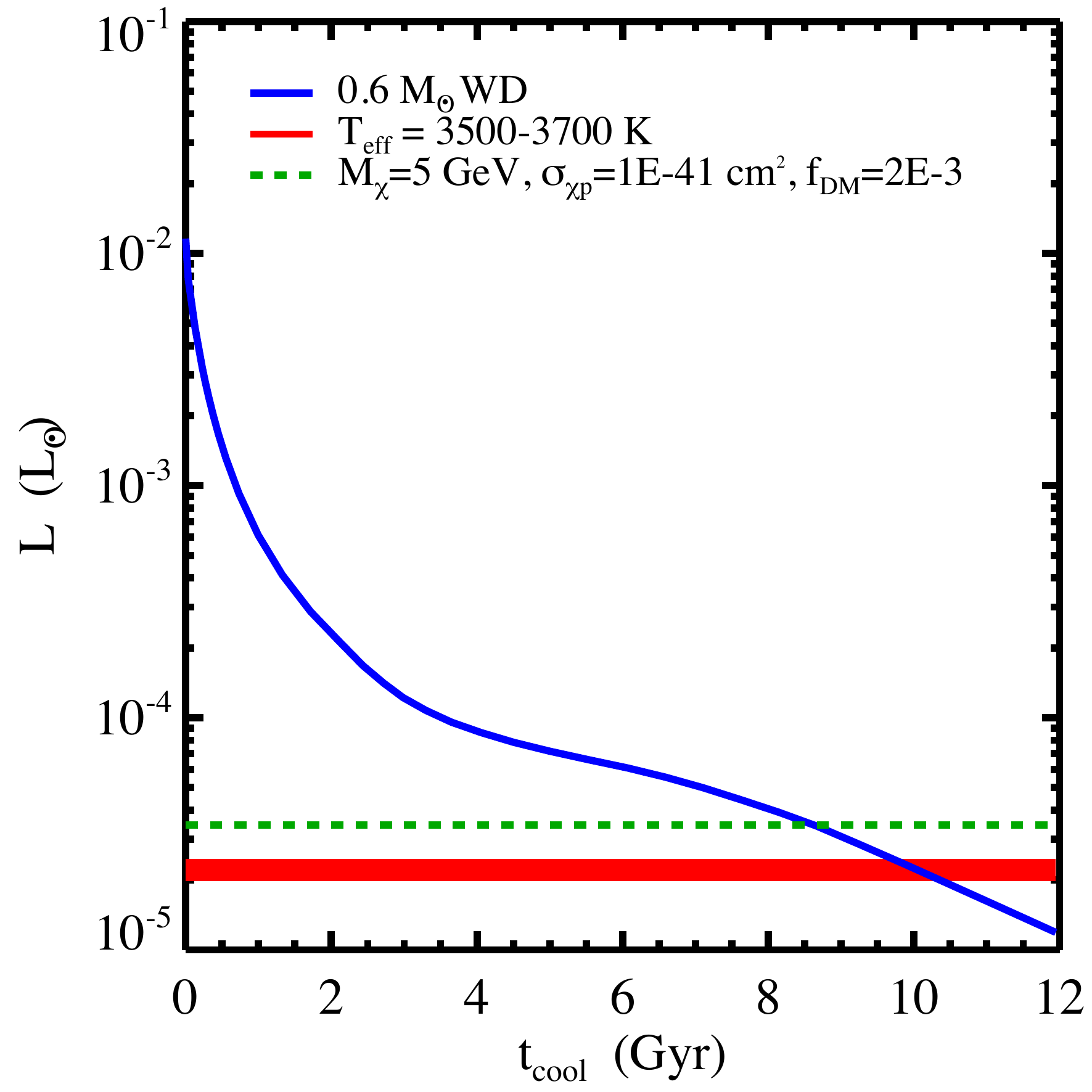}
\caption{
The luminosity $L$ in units of L$_\odot$ vs.\ cooling time $\tcool$.  The solid blue line represents the cooling curve of a 0.6 M$_\odot$ WD modeled with \MESAstar. The red band is the luminosity of a WD with $T_\mathrm{eff} \sim$ 3500-3700~K.  This is a similar temperature to that of the coolest WDs in NGC 6397, which we use to produce our constraints.  Note that  for a WD to reach such a temperature, $\tcool$ must be of order the age of the Galaxy.  The dashed green line represents the predicted luminosity from DM annihilations for a model with a DM fraction in NGC 6397 $\fdm = 2\, \times\, 10^{-3}$ and a DM particle with $\Mx = 5\, \gev$ and $\sigmaxp = 1.0\, \times\, 10^{-41}\, \cm^2$.  Such a model is consistent with the bounds from LUX and SuperCDMS while lying near to the confidence regions for the other experiments (see Fig.~\ref{fig:mvsigma}).  This model is clearly excluded by the data as the luminosity from DM annihilations must be less than the observed luminosity of the WD.  This demonstrates the potential for WIMP astronomy to constrain the density of DM in globular clusters as the assumed DM fraction is $\sim~3$ orders of magnitude below the constraint of \cite{Shin}. 
}
\label{fig:WDcool}
\end{figure}

The data we exploit in order to explore possible constraints on dark matter and/or globular cluster properties are the deep observations of the globular cluster NGC 6397 with the Hubble Space Telescope (HST) \cite{Heyl,Stello,DeMarchi,Richer}. In particular, we use observations of the WD cooling sequence \cite{Hansen}.  The WDs observed in NGC 6397 are among the coolest, dimmest WDs that have been observed to date and, importantly, the authors provide compelling evidence that they have observed the limit of the WD cooling sequence in NGC 6397. GCs, such as NGC 6397, are interesting objects in and of themselves in the context of WIMP astronomy. Globular clusters are the largest structures known that show no evidence of DM. Moreover, while the prevailing view is that GCs formed without significant DM halos, the viability of specific formation and evolution scenarios for the Galactic GCs remains controversial. The controversy is based, in part, around the complex element abundance patterns measured among the stars in several GCs \cite{Gratton,Gratton2012,Con&Sperg} which suggest that the clusters must have had significantly more mass in the past in order to retain significant amounts of heavy elements ejected from supernovae. Meanwhile, state-of-the-art constraints on the contemporary amounts of DM in GCs are not particularly restrictive. Ref.~\cite{Conroy} found that for NGC 2419 and MGC1 $M_{\mathrm{DM}}/M_* \lesssim 1$, where $M_{\mathrm{DM}}$ is the mass in DM of the cluster and $M_*$ is the stellar mass of the cluster. Subsequently, Ref.~\cite{Shin} found a similar result for NGC 6397. In the present work we show how WIMP astronomy may be able to improve upon these types of constraints by several orders of magnitude. We will show that if any of the anomalous signals observed in direct search experiments could be convincingly attributed to DM, the corresponding constraint on the density of DM in NGC 6397 would be $M_{\mathrm{DM}}/M_* \lesssim 10^{-3}$, at least three orders of magnitude more restrictive than the kinematic constraints.

The remainder of the paper is organized as follows:  In \S\ref{section:methods} we demonstrate how observations of cool WDs can be used to constrain DM particle and halo model parameters in 4 parts.  We first calculate the capture rate of WIMPs in the star in \S\ref{section:capture} .  We then calculate the annihilation luminosity in \S\ref{section:annihilation}.  Next, in \S\ref{section:WD}, we consider the effect of the annihilation luminosity on WD cooling.  Finally, in \S\ref{section:gc}, we use the cooling sequence of NGC 6397 to relate the temperature of the coolest WD observed to the annihilation luminosity.  We present our results in \S\ref{section:results} and summarize our conclusions and discuss possibilities for future applications of our method in \S\ref{section:conclusion}.

\section{Methods}
\label{section:methods}

In this section, we show how observations of cool WDs in GCs may lead to constraints on the DM particle and/or the dark matter content of GCs. We first discuss the rates of capture and annihilation of WIMPs in WDs.  We then show that if the annihilation luminosity is to be comparable to the luminosity of the WD, an equilibrium may have been reached between DM capture and annihilation.  Equilibration is achieved for a wide range of interesting parameter space, yielding a relationship between the DM particle properties, the DM environment of the WD and the observed luminosity of the WD.  The annihilation luminosity is particularly relevant in the case of WDs because their cooling will be halted by the injection of energy in the core from DM annihilations.  Therefore, the assumption that the luminosity of the WD is due to DM annihilations alone yields an upper limit on the local DM density. In the present work we use observations of the WD cooling sequence in NGC 6397 to derive constraints on the DM density of the cluster and DM model parameters $\Mx$ and $\sigmaxp$.

\subsection{The Capture Rate of WIMPs}
\label{section:capture}

The capture of DM particles in stars has been studied in numerous papers (e.g \cite{Griest&Seckel,GouldCap1,GouldCap2,Press&Spergel,Krauss,Faulkner}).  The rate at which a WD will capture DM is approximately given by Eq.~(A.16) from Ref.~\cite{Zentner}, namely 
%
\begin{equation}
\label{eq:capture_rate}
	C_{\mathrm{c}} = \sqrt {\frac{3}{2}}\frac {\rho_\chi}{\Mx}\sigma_i v_{\mathrm{esc}}(R)\frac{v_{\mathrm{esc}}(R)}{\bar{v}}N_i \langle\phihat\rangle\frac{\mathrm{erf}(\eta)}{\eta},
\end{equation}
where $\rho_\chi$ is the local DM density, $R$ is the radius of the WD, $v_{\mathrm{esc}}(R)$ is the escape speed at the surface of the star, $\bar{v}$ is the dispersion of the DM velocity profile (assumed to be Maxwell-Boltzmann), $\eta$ is the ratio of the star's velocity through its halo to the local velocity dispersion of the halo (assumed to be of order unity) and 
%
\begin{equation}
	\phihat = \frac{v_\mathrm{esc}^2(r)}{v_\mathrm{esc}^2(R)}
\end{equation}
is a dimensionless potential for stellar nucleons.  The quantity $\langle\phihat\rangle$ is the average of $\phihat$ over all nucleons in the star.  The cross-section for scattering off of nuclear species $i$ is $\sigma_i$, $N_i$ is the number of nucleons of species $i$ in the star, and the total capture rate is the sum over all species.  
%
Note that Eq.~(\ref{eq:capture_rate}) relies upon the assumption that the elemental abundance is not a function of the radius.  This is a reasonable approximation as the WD is the remaining core of a dead star, the abundance should not be a strong function of the radius.  Moreover, almost all of the interior is carbon and oxygen, which have similar scattering cross-sections and therefore yield virtually identical constraints. 

Calculation of $\langle\phihat\rangle$ requires that we know the density profile of the star.  We use the publicly availble stellar evolution code $\MESAstar$~\cite{MESA1,MESA2,MESAsite} to find the density profile and other relevant parameters we require.
%
%
The specific model we ran of a 0.6 M$_\odot$ WD is provided in the $\MESA$ test suite in the folder $\texttt{wd\_cool\_0.6M}$.  With this density profile we find $\langle\phihat\rangle \simeq 2.4$.

  Scattering off a given nucleus is coherent at most of the relevant energies that we consider, so the cross-section for scattering off a given nucleus is approximately 
%
\begin{equation}
	\sigma_N \approx \sigmaxp A_N^2\frac{\Mx^2 M_N^2}   {{(\Mx+M_N)}^2}\frac{{(\Mx+m_{\mathrm{p}})}^2}{\Mx^2m_{\mathrm{p}}^2},
\label{xsection}
\end{equation}
where $A_N$ is the atomic mass number of the nucleus of interest, $M_N$ is the mass of the nucleus, and $m_{\mathrm{p}}$ is the proton mass.  Capture of WIMPs with $\Mx \gtrsim 10\, \gev$ is modestly suppressed compared to Eq.~(\ref{xsection}) due to loss of coherence.  To account for this, we apply an exponential form-factor suppression as in \cite{GouldCap1,Jungman}
%
\begin{equation}
	f(Q) = \exp(-Q/2Q_0)
\end{equation}
where $Q$ is the energy transferred from the WIMP to the nucleus and $Q_0 = \frac{3}{2M_N R_N^2}$ is the `coherence energy' and $R_N = 10^{-13}\, \cm \times [0.91(M_N/\gev)^{1/3}+0.3]$ is the nuclear radius.

For simplicity we consider a WD made entirely from carbon and oxygen. This is a reasonable simplification because the progenitors of low mass WDs should not have been hot enough to fuse carbon in their cores and helium is only abundant in the thin stellar atmosphere and thus is unimportant for computing the capture rate.  Additionally, the GC we consider below is a metal-poor cluster \cite{Harris} so we do not expect a significant contribution from metals other than carbon and oxygen.  Thus we can write the composition of the WD as 
%
\begin{equation}
	N = N_{\mathrm{C}}  + N_{\mathrm{O}} = f_{\mathrm{C}}\frac{M_{\mathrm{WD}}}{M_{\mathrm{C}}} + (1 - f_{\mathrm{C}})\frac{M_{\mathrm{WD}}}{M_{\mathrm{O}}},
\end{equation}
where $f_{\mathrm{C}}$ is the fraction of Carbon.  Our results are not sensitive to the composition of the WD.  
%
\begin{figure*}[htp]
\centering
\subfloat[The color-magnitude diagram of NGC 6397.]{\includegraphics[width=9cm, height=9cm]{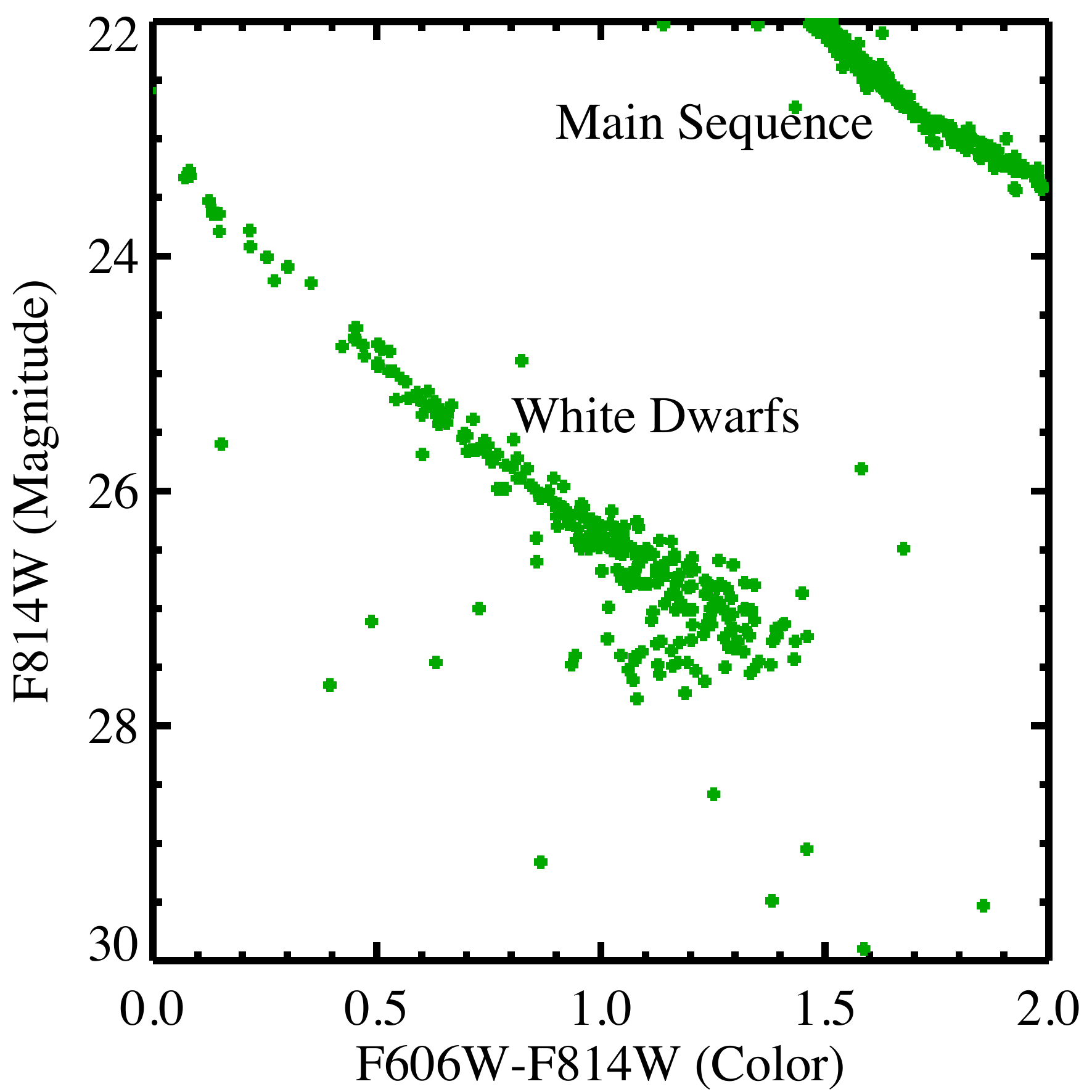}}
\subfloat[The WD luminosity function of NGC 6397.]{\includegraphics[width=9cm, height=9cm]{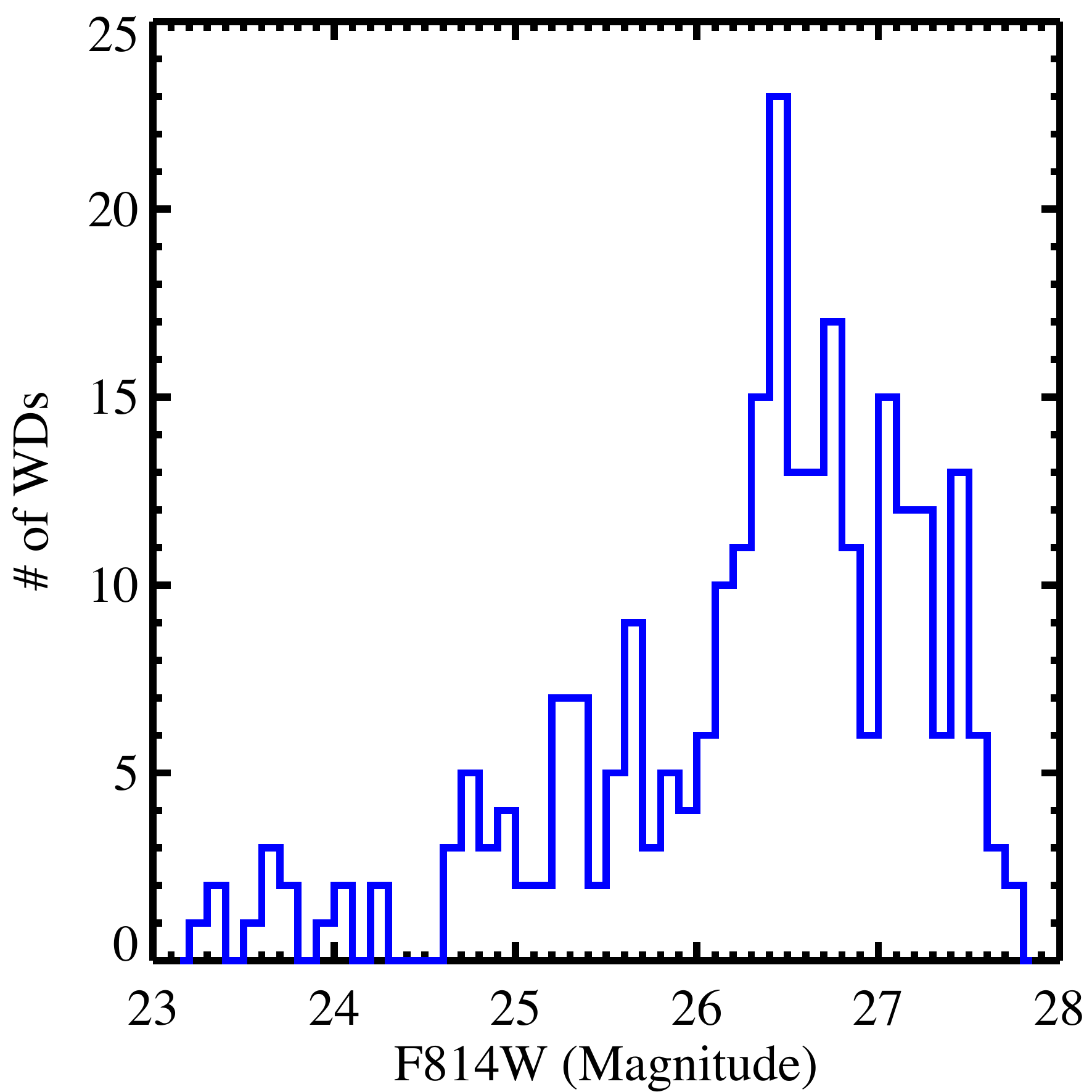}}
\caption{Left panel: The color-magnitude diagram of NGC 6397 shows a clear WD cooling sequence \cite{Hansen,Richer}.  The important feature is the sharp decline in the number of WDs with apparent magnitudes greater than F814W = 27.6.  The objects with magnitudes greater than this are unresolved galaxies.  The fact that these fainter objects are observed indicates that there is a real truncation of the WD cooling sequence.  The Main Sequence can be seen in the upper-right corner. Right panel:  The luminosity function of WDs in NGC 6397. Again, note the truncation at F814W = 27.6. The data points for both plots were obtained from Ref.~\cite{CDS}.
}
\label{fig:WDplot}
\end{figure*}
%
%

%
%

At sufficiently high scattering cross-sections $\sigmaxp~>~\sigmaxp^{\mathrm{crit}}$ (see Fig.~\ref{fig:mvsigma} below), the WD will become opaque to the WIMPs it encounters at which point DM capture will be saturated.  The critical cross-section can be evaluated by requiring that $\sigman^\mathrm{crit}\sim\Sigma_N^{-1}$, where $\Sigma_N = \frac{M_\mathrm{WD}/M_N}{\pi R_\mathrm{WD}^2}$ is the average projected surface density of the WD seen by infalling WIMPs.  This implies that 
%
\begin{equation}
	\sigmaxp^\mathrm{crit} = \frac{\sigman}{A_N^2}\frac{(\Mx+M_N)^2}{M_N^2}\frac{m_\mathrm{p}^2}{(\Mx+m_\mathrm{p})^2}.
\end{equation}
We assume that scattering cross-sections above this threshold all give the same prediction as $\sigmaxp^\mathrm{crit}$.

For scattering cross-sections below the saturation threshold, Eq.~(\ref{eq:capture_rate}) will be relevant in the limit that nearly all scattering events lead to capture within the star. The probability for capture in an individual DM-nucleon encounter is determined by the kinematics of the scattering and, assuming isotropic scattering, is given by (see the Appendix of Ref.~\cite{Zentner})
%
\begin{equation}
\label{eq:probability}
	P = \frac{v_{\mathrm{esc}}^2}{v_{\mathrm{esc}}^2 + u^2}\, \left[1 - \frac{u^2}{v_{\mathrm{esc}}^2} \, \frac{(\Mx-M_N)^2}{4\Mx M_N}\right],
\end{equation}

where $v_{\mathrm{esc}}$ is the escape speed at the position of the scattering event, $u$ is the speed at infinity of the incident DM particle, and $M_N$ is the mass of the nucleon involved in the scattering. Capture is extremely likely in nearly all cases of interest because $u \sim \bar{v} \ll v_{\mathrm{esc}}$. In the case of a $0.6\, \msolar$ WD with $R \sim 0.012\, \mathrm{R}_\odot$, which we consider below, $v_{\mathrm{esc}}(R) \approx 4400\, \mathrm{km/s}$ and the relevant escape speed is $v_{\mathrm{esc}}^2 \sim \langle\phihat\rangle v_{\mathrm{esc}}^2(R)$. Typical relative velocities are $u \sim \bar{v} \approx 10\, \mathrm{km/s}$ within GCs and $u \sim \bar{v} \approx 270\, \mathrm{km/s}$ in the Milky Way halo. The first factor in Eq.~(\ref{eq:probability}) differs from unity by $\lesssim 1\%$ in both the GC and Milky Way environments.

Consequently, capture will be efficient so long as the mass of the dark matter particle and target nucleon are not too different. Requiring the second term within brackets in Eq.~(\ref{eq:probability}) to be negligible leads to a mass range over which this approximate capture rate is viable, 
%
%
%
\begin{equation}
\label{eq:massrange}
	\frac{u^2}{v_{\mathrm{esc}}^2} \, \frac{M_N}{4} \ll \Mx \ll 4 \, \frac{v_{\mathrm{esc}}^2}{u^2}\, M_N.
\end{equation}
Assuming that $M_N \simeq 11.2\, \gev$, as would be the case for Carbon, this leads to an approximate dark matter particle mass range of $6~\mathrm{keV}~\ll~\Mx~\ll~2~\times~10^4~\mathrm{TeV}$  $(4~\mathrm{MeV}~\ll~\Mx~\ll~29~\mathrm{TeV}$) in the GC (Milky Way) environment. Outside of this mass range, the capture rate is suppressed significantly compared to Eq.~(\ref{eq:capture_rate}). In nearly all cases, other considerations will limit the mass range over which our calculations are relevant. For example, as we discuss below, dark matter particles with masses $\Mx \lesssim 100\, \mathrm{keV}$ evaporate from WDs more rapidly than they are captured, so there is no accumulation of low-mass dark matter in WD stars. Meanwhile, loss of coherence in the scattering event leads to modest form-factor suppression for $\Mx \gtrsim 10\, \gev$, and the upper range of $\Mx$ in the GC environment exceeds the unitarity bound on thermal relic dark matter \cite{griest_kamionkowski90}.

In fact, while Eqs.~(\ref{eq:capture_rate})~\&~(\ref{xsection}) are useful approximations to guide the reader, when calculating the capture rate we actually use the full results of the appendix in Ref.~\cite{GouldCap1}, which account for the exponential form-factor suppression.

\subsection{The Annihilation Luminosity}
\label{section:annihilation}

An important aspect of our constraints is that they extend to much lower masses than most other techniques, Cosmic Microwave Background (CMB) constraints being the notable exception \cite{pwave, Diamanti, Aravind} (these constraints are discussed at the end of \S\ref{section:results}).  We consider stable DM of mass $\Mx \gtrsim$ 100 keV.  WIMPs with $\Mx \lesssim$ 100 keV can escape the WD via evaporation---the ejection of WIMPs by hard elastic scattering from nuclei \cite{Jungman}. We can estimate the evaporation mass by demanding that the typical velocity of a WIMP $v \sim (T_\mathrm{c}/\Mx)^{1/2}$ be less than the local escape speed of the star $\vesc\approx 1.1\, \times 10^4$ km/s. Here $T_\mathrm{c}$ is the temperature in the core of the star \cite{Jungman}. Below we will consider a WD with $M_\mathrm{WD} \approx 0.6\, \mathrm{M}_\odot$ and $T_\mathrm{eff} \approx$ 3700 K. Using \MESAstar, we find that such a WD should have a core temperature $T_\mathrm{c} \sim 10^6$ K corresponding to an evaporation mass $\sim$ 100 keV.  (For a more careful discussion of the evaporation mass see \cite{Griest&Seckel}).  Thus, we are justified in neglecting evaporation for $\Mx \gg 100\, \kev$.

Having computed the capture rate of DM and justified the neglect of DM evaporation, the number of WIMPs contained within the star, $N_\chi$, is governed by the differential equation 
%
\begin{equation}  
	\frac{\mathrm{d}N_\chi}{\mathrm{d}t} = C_{\mathrm{c}} - C_{\mathrm{a}}N_\chi^2,
\end{equation}
where $C_{\mathrm{a}}$ is twice the rate of annihilation events (because each annihilation eliminates 2 particles). The solution to this equation for homogeneous initial conditions is
%
\begin{equation}
	N_\chi = \sqrt{\frac{C_{\mathrm{c}}}{C_{\mathrm{a}}}}\tanh\left(\sqrt{C_{\mathrm{c}}C_{\mathrm{a}}}t\right).
\end{equation}

There is a timescale for equilibration between DM annihilation and capture, $\tau_{\mathrm{eq}} = 1/\sqrt{C_{\mathrm{c}}C_{\mathrm{a}}}$, such that for $t \gg \tau_{\mathrm{eq}}, N_\chi$ approaches a steady state solution $N_{\chi,\mathrm{eq}} = \sqrt{C_{\mathrm{c}}/C_{\mathrm{a}}}$ \cite{Zentner}.  The annihilation rate at equilibrium within a WD will be
%
\begin{equation}
	\Gamma_{\mathrm{a}} = \frac{1}{2}C_{\mathrm{a}}N_{\chi,\mathrm{eq}}^2 = \frac{1}{2}C_{\mathrm{c}},
\end{equation}
because there are $N_{\chi,\mathrm{eq}}^2$/2 distinct pairs of DM particles within the star.  To calculate $\tau_{\mathrm{eq}}$ we first express $C_{\mathrm{a}}$ in terms of effective volumes \cite{Griest&Seckel}
%
\begin{equation}
	C_{\mathrm{a}} = \langle\sigma_{\mathrm{a}}v\rangle\frac{V_2}{V_1^2},
\end{equation}
where $\langle\sigma_{\mathrm{a}}v\rangle$ is the thermally averaged annihilation cross-section, and 
%
\begin{equation}
	V_j \simeq 9.1\times 10^{21}\left(\frac{1\ \gev}{j\Mx}\right)^{3/2}\cm^3
\label{eq:v_eff}
\end{equation}
is the effective volume of captured DM particles within the star for $j$ = 1 \cite{Griest&Seckel,Zentner}.  Within the volume of interest the density is approximately $\rho(r) \approx \rho_\mathrm{c} = 4\, \times\, 10^6$ g/cm$^3$ and the temperature is approximately $T(r) \approx T_\mathrm{c} = 10^6\, \mathrm{K}$, where the subscript c denotes the value in the core and both $\rho_\mathrm{c}$ and $T_\mathrm{c}$ were estimated using \MESAstar.  Note that when $\Mx \lesssim 10^{-2}\, \gev$ the effective volume will be $\sim 20\%$ of the volume of the WD.  At this point our approximations will begin to break down as it is not the case that $\rho(r) \approx \rho_\mathrm{c}$.  However, our results are robust to this effect, because we are in the equilibrated regime by many orders of magnitude (see the following paragraph) and the annihilation rate can therefore be determined by the capture rate of WIMPs alone.  
%
\begin{figure}[htp]
\centering
\includegraphics[width=9cm, height=9cm]{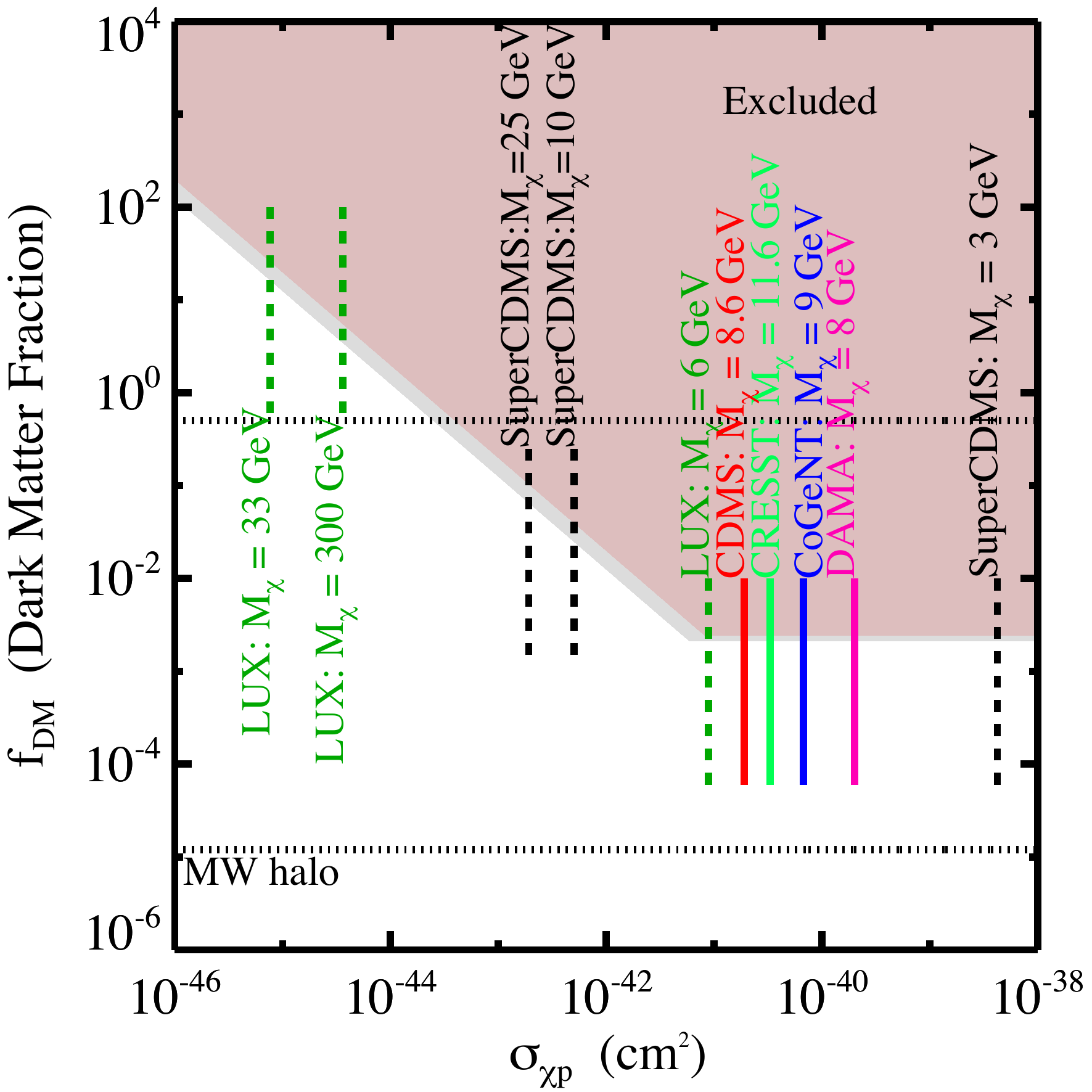}
\caption{
The DM fraction vs. DM-proton scattering cross-section for NGC 6397. The grey shaded band represents the constraints for the best fit values of $\Mx$ from the direct detection experiments and spans the range 3500-3700 K corresponding to the temperature of the WD at the truncation of the cooling sequence.  (The constraints from CoGeNT, CRESST, DAMA and CDMS all lie within this band).  The reason for the flattening of the constraint at $\sigmaxp \sim 10^{-41} \cm^2$ is that $\sigmaxp > \sigmaCrit$ and capture is saturated.  The pinkish shaded region is the excluded portion of parameter space.   The horizontal dotted lines show the fiducial contraints on  $\fdm$ (see \S\ref{section:gc}). The solid vertical lines denote the best fit cross-sections for CoGeNT, CRESST, DAMA and CDMS as indicated by the labels on the plot while the vertical dashed lines indicate the exclusion limits from LUX and SuperCDMS at several different values of $\Mx$ as indicated by the labels.  Note that if any of the scattering events from CoGeNT, CRESST, DAMA and CDMS are confirmed as DM WIMPs, then the fraction of DM in NGC 6397 must be at least 2 orders of magnitude below the fiducial upper limit, and is within 3 orders of magnitude of the ambient Galactic DM density.
}
\label{fig:rhovsigma}
\end{figure}
%
%

%
%

For a canonical, thermal WIMP DM particle with $\langle\sigma_{\mathrm{a}}v\rangle = 3\, \times\, 10^{-26}\, \cm^3/\mathrm{s}, \sigmaxp = 10^{-41}\, \cm^2$ and mass $\Mx = 10$ GeV, captured by a WD like that at the truncation of the cooling sequence in NGC 6397, we find that $\tau_{\mathrm{eq}}$ is less than a year for the fiducial constraint with $\rho_\chi = \rho_*$. For our more conservative constraints $\rho_\chi \lesssim \rho_*\times\,10^{-3}$ (dervied below) we find that $\tau_{\mathrm{eq}} \gtrsim 1$ year. In either case, this is many orders of magnitude less than the age of the cluster $\sim 10$~Gyr. Clearly, the equilibrium relation for the annihilation rate within the star is valid for a large swath of interesting parameter space.

In fact, $\tau_{\mathrm{eq}}$ is of order the age of the cluster only for exceptionally small annihilation rates on the order of $\sigmav\lesssim 10^{-49}\, \cm^3/\mathrm{s}$ for the fiducial constraint above and $\sigmav \lesssim 10^{-46}\, \cm^3/\mathrm{s}$ for our more conservative constraints with $\rho_\chi \lesssim \rho_*\times\, 10^{-3}$. An important consequence of this fact is that this method can be used to probe dark matter with annihilation cross-sections many orders of magnitude smaller than the canonical WIMP cross-sections. Moreover, our results are insensitive to the DM annihilation cross-section over many orders of magnitude.

For a specific DM model, DM annihilations give rise to an energy source within the WD of luminosity 
%
\begin{equation}
\label{eq:Lchi}
	L_\chi \approx \Gamma_{\mathrm{a}}\Mx.
\end{equation}
Interestingly, $\Gamma_{\mathrm{a}} \propto \Mx^{-1}$ in the limit that the kinematics are favorable for DM capture [Eq.~(\ref{eq:capture_rate})], so this luminosity is independent of the DM particle mass over many orders of magnitude. Note that this is not merely a coincidence, but is a consequence of the fact that the flow of DM mass through the star is independent of the WIMP mass and the star is equally effective at capturing mass from this flow over many orders of magnitude.

If the annihilations proceed to SM particles, nearly all of the energy released at the annihilation will be deposited in the interior of the star, as the only particles capable of escaping the star are low energy neutrinos.  To show this, we first note that the projected surface density seen by neutrinos leaving the core is $\Sigma_N = \int^R_0{\rho_N(r)\, \mathrm{dr}}.$  If we assume for the moment that the WD is made entirely of carbon, then $\Sigma_\mathrm{C} = 6.8\, \times\, 10^{37}\, \cm^{-2}$ (using the density profile from \MESAstar).  The WD will be opaque so long as the neutrino-nucleus scattering cross-section is greater than $\Sigma_\mathrm{C}^{-1} = 1.4\, \times\, 10^{-38}\, \cm^2$.  Therefore, the WD will be opaque to neutrinos with energies $E_\nu \gtrsim 1\, \gev$ (see e.g. Fig.~9 in Ref.~\cite{Formaggio}).  Thus, models of low mass DM annihilating directly to neutrinos are capable of evading our constraints.

\subsection{Deriving the Constraint from a Cool White Dwarf}
\label{section:WD}

WDs cool and dim over time (see Figs.~\ref{fig:WDcool}~\&~\ref{fig:WDplot} and Ref~\cite{HansenReview}). If DM annihilations occur at a significant rate within a WD, then annihilations constitute a source of energy within the WD that can eventually provide an amount of energy comparable to the luminosity of the WD. If this occurs, the WD cooling will be halted and the WD will come to equilibrium at a minimum luminosity, $L \approx L_\chi$ and corresponding minimum effective surface temperature. This surface temperature can be related to a bolometric luminosity using a stellar model, which we calculate using \MESAstar.  Consequently, the observed luminosities and surface temperatures of WDs can be used to place bounds on DM properties and the distribution of DM local to the WDs. (Though we do not discuss them here, another well motivated DM particle candidate is the axion.  It is worth noting that WDs are also used to constrain the parameters of axions, as they would also have observable effects on WD cooling \cite{Raffelt,Isern}.)

Fig.~\ref{fig:WDcool} shows the cooling curve of a 0.6~$M_\odot$ WD as a function of the cooling time $\tcool$ as modeled in \MESAstar.  Also shown are the luminosity of the coolest WD in NGC 6397, which we use to derive our constraints, and the annihilation luminosity for a DM particle with $\Mx = 5\, \gev$ and $\sigmaxp = 1.0\, \times\, 10^{-41}\, \cm^2$ and an assumed DM density in NGC 6397 that is $2\, \times\, 10^{-3}$ times the stellar density.  The annihilation luminosity is well above the observed luminosity of the WD; therefore, this model is ruled out by the data.
%
\begin{figure*}[htp]
\centering
{\includegraphics[width=18cm, height=9cm]{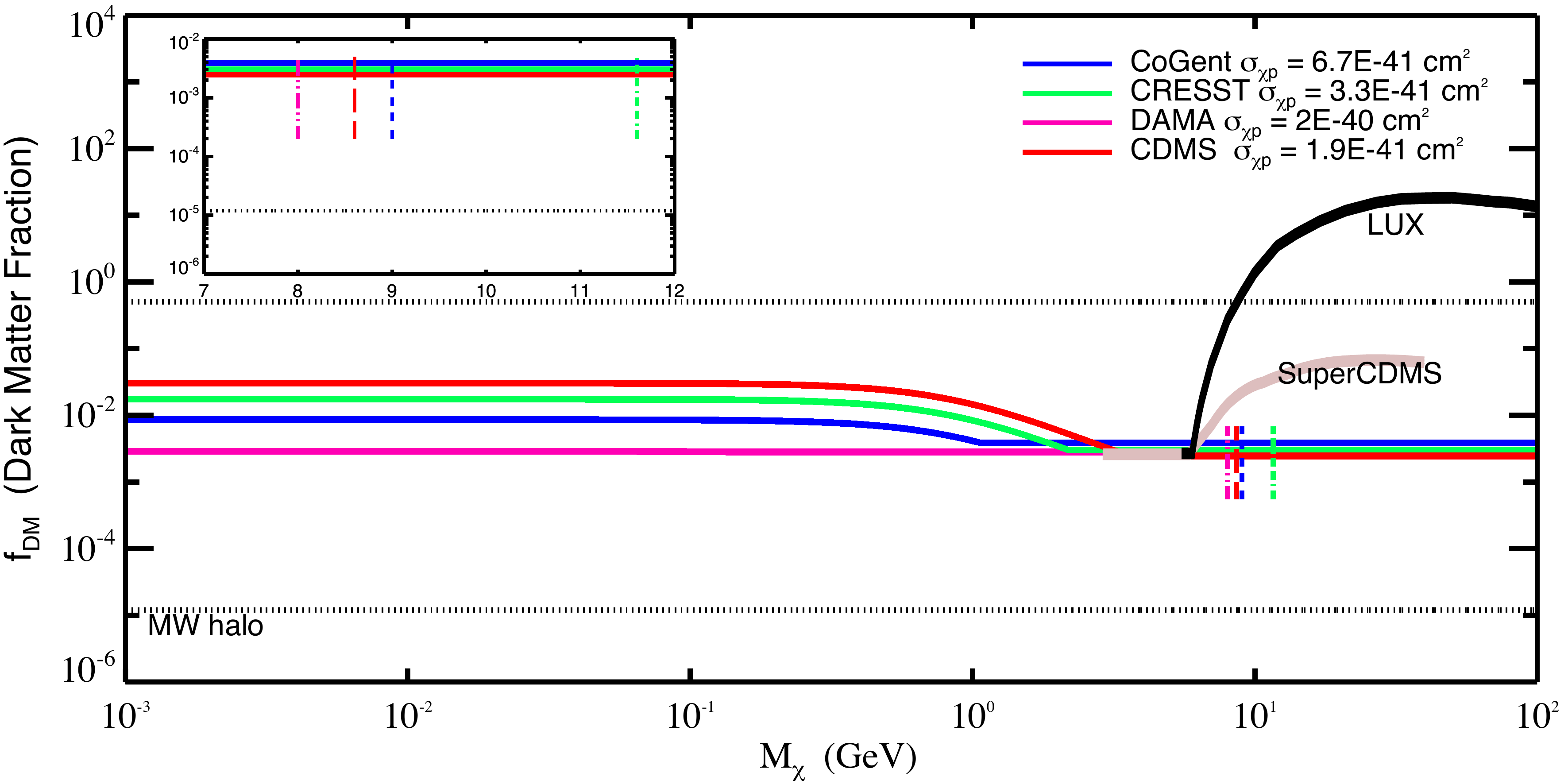}}
\caption{
The DM fraction vs. DM particle mass for NGC 6397. The colored bands are the constraints for the best fit values of $\sigmaxp$ from CDMS-Si, CRESST, CoGeNT, and DAMA (from top to bottom). The widths correspond to the uncertainty in the luminosity of the dimmest WDs used to observe the truncation of the WD cooling sequence in NGC 6397.  The horizontal dotted lines are the fiducial constraints on $\fdm$.  The vertical dashed lines denote the best fit mass for the corresponding experiment. Also shown are the 90\% exclusion limits from LUX and SuperCDMS \cite{DMTools}.  The constraints all flatten out when $\sigmaxp$ exceeds $\sigmaCrit$ and capture is saturated.  These constraints place an upper limit $\fdm \lesssim 10^{-2}$ $independent\ of \Mx$.  It is important to note that our constraints extend to WIMP masses $\sim 1\, \mev$, demonstrating the unique power of this technique to constrain low mass WIMPs.
}
\label{fig:rhovmass}
\end{figure*}
%
%

%
%

This is the effect that we aim to exploit. In a particular astrophysical environment (we will explore GCs) DM particles may annihilate within WDs providing a source of annihilation luminosity that is proportional to the DM-nucleon scattering cross-section $\sigmaxp$ and the local DM density $\rho_\chi$, and inversely proportional to the local DM velocity dispersion $\bar{v}$. Typical DM velocities must be similar to the local stellar velocities for an equilibrium structure, so $\bar{v}$ is informed by observational data and has little parametric freedom compared to the DM density and cross-section. According to this argument, the minimum luminosity of WDs in a given environment constrains the product $\rho_\chi \sigmaxp$ such that $L \gtrsim L_\chi$.

WDs are particularly promising probes for WIMP astronomy because they have deeper potential wells, cooler interiors, and significantly lower luminosities than their Main Sequence counterparts. For these reasons, WDs can be used for WIMP astronomy even for dark matter candidates with masses far below those typically considered ($\Mx \ll 1\, \gev$) and with annihilation cross-sections orders of magnitude below the standard thermal value. These are important points that appear to have been missed in the discussions of Refs.~\cite{Hooper, Bertone}.
%
%

In the following section, we will present constraints on the DM density and scattering cross-section based on high-quality observations of a nearby GC.

\subsection{The Globular Cluster NGC 6397 and the Milky Way Halo}
\label{section:gc}

The WD cooling sequence of NGC 6397 has been measured to unprecedented depth and precision by the Advanced Camera for Surveys (ACS) aboard HST \cite{Hansen,Richer}, making it an ideal candidate for constraining dark matter and/or GC structure using WD cooling. NGC 6397 is a metal-poor GC located at a distance of 2.6 kpc from the Sun (making it one of the 2 closest GCs along with M4) \cite{Harris,Reid}. This cluster has a mass $M_{6397} = \left(1.1\ \pm\ 0.1\right) \times 10^5$ M$_\odot$ \cite{Heyl} determined from kinematics, half-light radius $R_{\mathrm{HL}}$ = 2.2 pc, velocity dispersion $\bar{v} = 4.5 \pm 0.2$ km/s, and is one of 29 Galactic GCs which has undergone a core-collapse \cite{Harris}.

The population of WDs in NGC 6397 has been measured well with deep HST imaging and the color-magnitude diagram of NGC 6397 exhibits a clear WD cooling sequence with a truncation at low luminosity and low temperature, as expected due to the finite age of the cluster \cite{Hansen}. The field in NGC 6397 observed by ACS lies 5' SE of the cluster center \cite{Richer} well beyond the tidal radius of this core-collapsed cluster ($r_{\mathrm{t}}$ = .6' \cite{Harris}) and ranges from just inside the half-light radius ($r_{\mathrm{HL}}$ = 174'' \cite{Richer}) to several arcminutes beyond $r_{\mathrm{HL}}$ such that the WDs mainly probe the outter regions of the cluster.  As discussed in Appendix A of Ref.~\cite{Hansen}, the truncation of the WD cooling sequence occurs at an absolute magnitude in the Hubble F814W filter $M_{814} = 15.15 \pm 0.15$. The best-fit model for the WD is a mass at truncation of roughly $M_{\mathrm{WD}} \simeq 0.6$ M$_\odot$ which corresponds to a cooling age of $t_{\mathrm{cool}} \simeq 11.0\, \pm 0.5\, \mathrm{Gyr}$. Using the cooling models of Bergeron \cite{Holberg,Kowalski,Bergeron,Fontaine,BergeronSite} and applying a correction for WD masses in the range 0.5-0.6 M$_\odot$, we find $T_{\mathrm{eff}} \simeq$ 3500-3700 K.

Fig.~\ref{fig:WDplot} shows the color magnitude diagram of WDs in NGC 6397 as imaged in the ACS filters F814W and F606W as well as the luminosity function of WDs in the cluster.  As mentioned above, the WD cooling sequence is seen quite clearly, as is the truncation at apparent magnitude F814W = 27.6.

In order to develop constraints on DM and/or the DM content of NGC 6397, it is necessary to understand the basic properties of the cluster and to parameterize the density of dark matter within the cluster. We estimate the average density of NGC 6397 as
%
\begin{equation}
\bar{\rho} \approx \frac{(M_{6397}/2)}{\frac{4}{3}\pi R_{\mathrm{HL}}^3} \approx 4.7 \times 10^4\ \mathrm{GeV/cm}^3,
\end{equation}
where we have assumed that half of the GC mass is enclosed by the half-light radius. Note that the mass measurement comes from the kinematics of the cluster; therefore, this average density includes any DM that may be present. Current bounds on the amount of DM are roughly at the level of $M_{\mathrm{DM}}/M_{6397} \lesssim 1$ \cite{Conroy,Shin}. We will demonstrate that if the properties of dark matter become known and reside within a large range of viable parameter space, observations of the WD cooling sequence will provide significantly more restrictive constraints on any DM component associated with GCs.

There is little guidance on the structures of DM halos that GCs may have had early in their evolution, therefore we parameterize the amount of DM in the cluster by the fraction of DM,
%
\begin{equation}
\fdm = \rho_\chi/\bar{\rho},
\end{equation}
and take $\rho_\chi$ to be a constant function of position within the half-light radius of the cluster. The aforementioned constraints on the dark matter contained within NGC 6397 \cite{Shin} limit $\fdm \lesssim 0.5$. Meanwhile, the ambient DM contributed by the halo of the MW galaxy places a lower limit on $\fdm$. We estimate the lower limit on $\fdm$ by scaling the present day local density of DM~$\sim~0.4~\gev/\cm^3$ as $1/r$.  Taking the Galactocentric distance of NGC 6397 to be $6.0\, \kpc$ \cite{Harris} the density in the vicinity of the cluster is $\rho_\chi~\sim0.57\, \gev/\cm^3$.
%
%

The density and surface brightness profiles of GCs are typically described by King models which have constant density cores \cite{King1962,King1966} (as a post core-collapse cluster, NGC 6397 has a more centrally-concentrated profile than the typical King model). This is in contrast to the density profiles of galaxy-scale DM halos, which increase like $1/r$ towards their centers \cite{NFW}. Thus, our parametrization of the DM fraction should be conservative in the sense that, if NGC 6397 has a DM halo with a steep density profile as expected within the context of CDM, then $\rho_\chi$ should be much greater near the center of the cluster than at the half-light radius.

\section{Results}
\label{section:results}

By demanding that the annihilation luminosity of DM within the WDs in NGC 6397 [Eq.~(\ref{eq:Lchi})] does not exceed the observed luminosities of the least luminous WDs in the cluster (Section~\ref{section:gc} above), it is possible to constrain the product $\fdm \, \sigmaxp$ as a function of particle mass $\Mx$, so long as the annihilation rate for the DM exceeds $\sigmav~\gtrsim~10^{-46}\, \cm^3/\mathrm{s}$.\footnote{The precise threshold in annihilation cross-section depends upon $\fdm$, $\sigmaxp$, and $\Mx$, but this is a good rough number to have in mind and is many orders of magnitude smaller than the cross-sections considered for more typical WIMP-like thermal relic DM.} We show that it is possible to constrain the DM contents of the GC NGC 6397 (and thus GC evolution) if the DM properties become well constrained in the future, provided that the scattering cross-section is not too small.
%
%

\begin{figure}[htp]
\centering
\includegraphics[width=9cm, height=9cm]{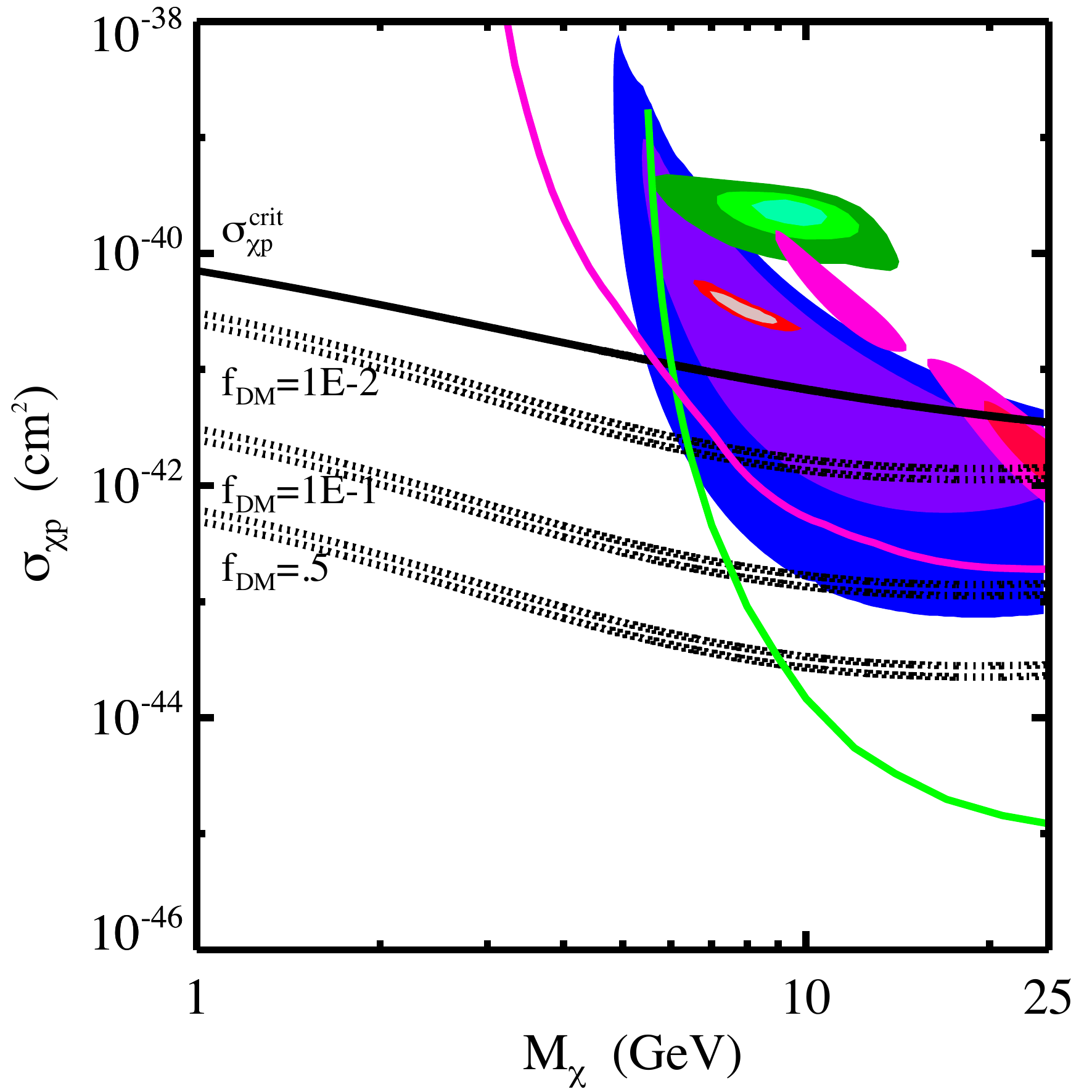}
\caption{
Constraints on the DM scattering cross-section $\sigmaxp$ as a function of mass $\Mx$ assuming various contributions to the DM content of the globular cluster NGC 6397. The black dotted curves correspond to the values of $\fdm$ indicated by the labels on the figure. The widths of the lines show the uncertainty in the constraints due to uncertainty in the precise luminosities of the dimmest WDs in NGC 6397. The solid black curve shows the critical cross-section $\sigmaCrit$ above which capture of DM particles is saturated, limiting our ability to constrain the DM WIMP parameters.  Shaded regions denote the confidence contours for CDMS-Si (95\% purple, 99\% blue), CRESST (1-sigma light red, 2-sigma pink), CoGeNT (90\% grey, 99\% dark red, near the upper left portion of the purple CDMS-Si contour), and DAMA (90\% cyan, 3-sigma light green, 5-sigma dark green). The solid curves represent the exclusion limits for LUX (95\% light green) and SuperCDMS (90\% pink). Confirmation of the anomalous scattering events as due to DM WIMPs has the potential to limit $\fdm \lesssim 10^{-3}$.  The limits from the various experiments were obtained from Ref.~\cite{DMTools}.
}
\label{fig:mvsigma}
\end{figure}

\begin{figure}[htp]
\centering
\includegraphics[width=9cm, height=9cm]{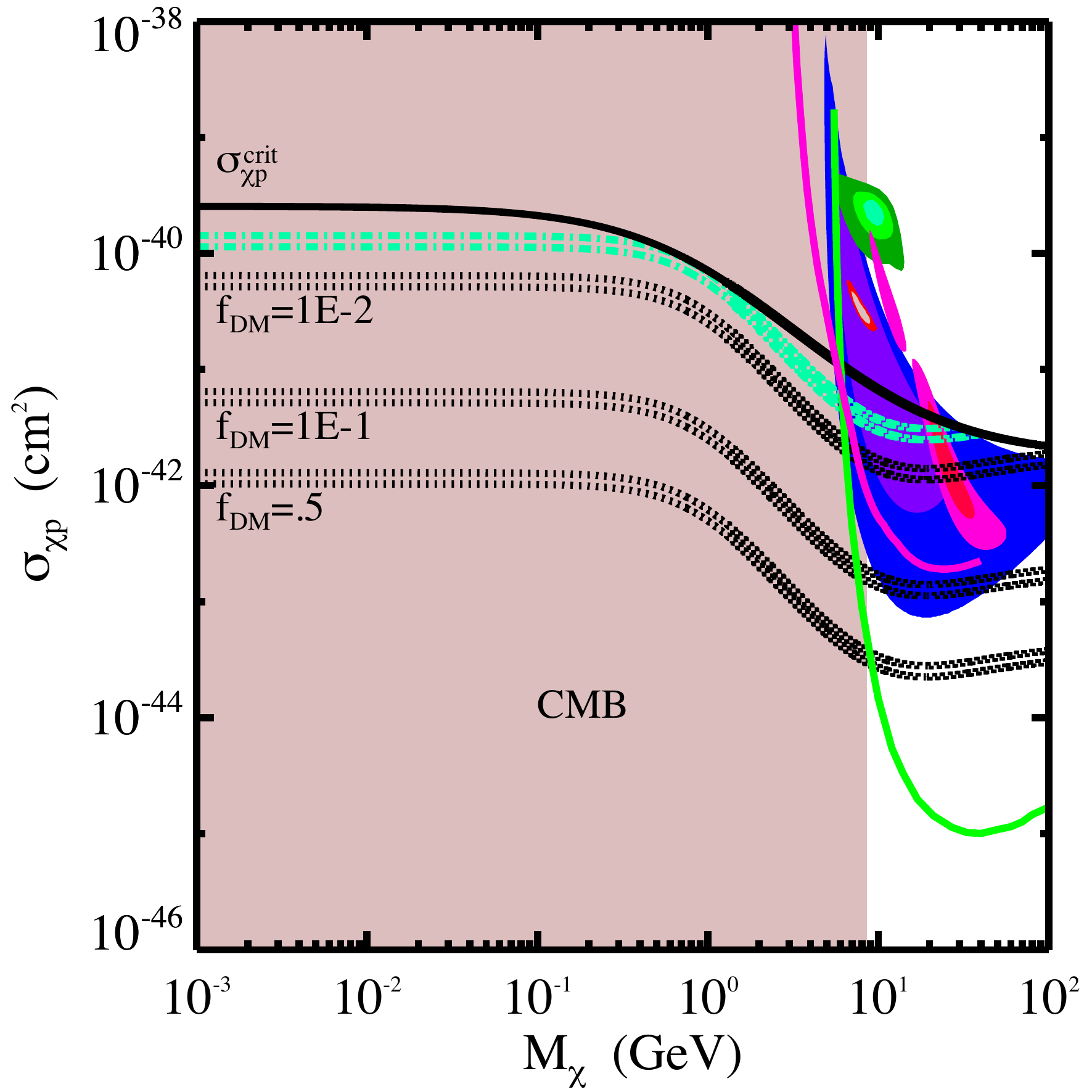}
\caption{
Constraints on the DM scattering cross-section $\sigmaxp$ as a function of mass $\Mx$ assuming various contributions to the DM content of the globular cluster NGC 6397. All curves and contours are the same as Fig.~\ref{fig:mvsigma} with the exception of the cyan dot-dashed curve, which shows the constraint from a hypothetical 3500-3700 K WD located in the dwarf galaxy Segue~I.  The pinkish shaded region is the region excluded by observations of the CMB under the assumptions of Ref.~\cite{Aravind}. Confirmation of the anomalous scattering events as due to DM WIMPs has the potential to limit $\fdm \lesssim 10^{-3}$.  As in Fig.~\ref{fig:rhovmass}, our constraints extend to much lower masses ($\sim 1\, \mev$) than those typically probed by other techniques ($\gtrsim 1\, \gev$).
}
\label{fig:mvsigma2}
\end{figure}

%
%
Constraints realized from WD cooling in GCs lie in the three-dimensional parameter space of $\Mx$, $\sigmaxp$, and $\fdm$, which we present in Figs.~\ref{fig:rhovsigma}-\ref{fig:mvsigma2}.  As an example of the astrophysical reach of WD cooling constraints, consider first constraints on $\fdm$ as a function of $\sigmaxp$.  These are depicted in Fig.~\ref{fig:rhovsigma} assuming the best-fit values of $\Mx$ from a DM interpretation of the recent CDMS-Si \cite{CDMS-Si}, CoGeNT \cite{CoGent}, CRESST \cite{CRESST}, and DAMA \cite{Savage} results. Fig.~\ref{fig:rhovsigma} shows that if such an interpretation of any of these results could be taken seriously, then it must be the case that either $\fdm \lesssim 10^{-3}$ or that the DM annihilates with a rate smaller than $\sigmav \lesssim 10^{-46} \, \cm^3/\mathrm{s}$, some {\em 20 orders of magnitude} lower than the canonical thermal relic WIMP value of $\langle \sigma_{\mathrm{a}}v\rangle \sim 3 \times 10^{-26} \cm^3$/s. 
%

Fig.~\ref{fig:rhovmass} shows the corresponding constraints for $\fdm$ as a function of $\Mx$ assuming the best-fit values of $\sigmaxp$ from a DM interpretation of the recent CDMS-Si \cite{CDMS-Si}, CoGeNT \cite{CoGent}, CRESST \cite{CRESST}, and DAMA \cite{Savage} results.  The consequences for $\fdm$ and the annihilation rate are the same as in Fig.~\ref{fig:rhovsigma}. On the other hand, DM parameter values that are not in conflict with the results of LUX and SuperCDMS can still place strong constraints on $\fdm$.  Fig. ~\ref{fig:rhovmass} also shows that if $\sigmaxp \sim 10^{-41}\,  \cm^2$, then $\fdm \lesssim 10^{-2}$ {\it independent of} $\Mx$.
%

Dark matter direct search experiments such as CDMS, CRESST, CoGeNT, DAMA, and LUX typically present their results in the plane of $\Mx$-$\sigmaxp$. Figs.~\ref{fig:mvsigma}~\&~\ref{fig:mvsigma2} show the constraints that WD cooling in NGC 6397 place on this parameter space for a variety of possible dark matter contributions $\fdm$. The confidence contours derived from the CoGeNT, CRESST, DAMA and CDMS data as well as the exclusion limits from LUX and SuperCDMS are also shown in Figs.~\ref{fig:mvsigma}~\&~\ref{fig:mvsigma2}.  In the mass range $\Mx \gtrsim 3\, \gev$, the WD cooling constraint is only competitive with direct search constraints from LUX and SuperCDMS if the DM fraction in NGC 6397 exceeds $\fdm \gtrsim 0.1$. While this value of $\fdm$ is not ruled out, there is no compelling reason to believe that NGC 6397 should have such significant DM content. In any case, the possibility of constraining $\fdm$ at this level is intriguing because it exceeds the potential of kinematic constraints.

The results shown in Figs.~\ref{fig:rhovsigma}-\ref{fig:mvsigma2} are interesting for several reasons. Interpreting any of the anomalous signals in the variety of experiments that have reported events in direct detection detectors as DM yields a constraint on the DM content of GCs that is far more restrictive than any kinematic constraints. Furthermore, even models that exhibit no tension with the recent LUX and SuperCDMS constraints may yield observably large effects on WD cooling so as to either be constrained by WD cooling in GCs or to constrain $\fdm$ within these GCs. This is particularly true at low DM particle masses ($\Mx \lesssim$ 2-3 GeV), where the WD cooling constraint can be more stringent than direct detection constraints.
%
%

Indeed, we follow-up on the results of Ref.~\cite{Hooper} by displaying the potential DM constraints that could be gleaned from the hypothetical observation of a similarly truncated WD cooling sequence in the dwarf galaxy Segue~I.  These constraints are interesting because unlike NGC 6397, Segue~I is known to possess DM.  We estimate the DM density and velocity dispersion in Segue~I to be 175 GeV/cm$^3$ and 3.7 km/s respectively \cite{Simon}.  The annihilation rates probed in Segue~I will then be $\sigmav \gtrsim 10^{-46}\, \cm^3$/s which is similar to those probed in NGC 6397.

%
%
Notice that these constraints may be uniquely restrictive in the low-mass regime, as constraints from particle accelerators tend to be strongly model-dependent.  A recent example is Ref.~\cite{Askew}, which explores constraints using an itemization of specific operators. Following the reasoning in this paper, it is clear that collider constraints are more significantly model dependent than our astrophysical bounds. Moreover, collider constraints do not extend to very low mass in general, largely because events are triggered on large missing energy which makes identification of low-mass dark matter difficult. An exception are constraints from LEP which can rule out light DM matter ($\Mx \lesssim 10\, \gev$) in certain scenarios \cite{LEP}.  However, these constraints are still model-dependent and more generally the constraints from LEP lie roughly 2-3 orders of magnitude above our constrains in the $\sigmaxp-\Mx$ plane (see the left panel in Fig.~3 of Ref.~\cite{LEP}). These are important distinctions. Meanwhile, the only competitive astrophysical constraints are those from observations of the CMB. Indeed, we estimate that WD heating can be exploited to constrain DM with masses as low as 100 keV.  Such low DM masses may be ruled out by the CMB.  For instance, Ref~\cite{Aravind} found that for s-wave annihilation to bottom quarks $\chi\chi\rightarrow b\bar{b}, \Mx > 8.6\, \gev$ at the 95\% confidence level while Ref.~\cite{pwave} found that for s-wave annihilation $\Mx >$ 5-26 GeV at the 2$\sigma$ level depending on the efficiency of energy injection.

However, CMB constraints on particle mass all rely on the assumption of a nearly standard s-wave thermal relic cross section. WD cooling can probe dark matter of both very low mass {\em and} very low annihilation rates down to $\sigmav \sim 10^{-46}\, \cm^3$/s in NGC 6397 and Segue I. Models with such low cross sections are not constrained by the CMB. One way to realize relatively low contemporary annihilation cross sections in a comprehensive model is through models in which s-wave annihilation is suppressed and p-wave annihilation dominates. CMB constraints on such scenarios are weak. As an example, Ref.~\cite{Diamanti} found that for $\Mx = 100$ MeV (1 GeV) $\sigmav \gtrsim 10^{-24} (10^{-23})$ cm$^3$/s at a reference speed of 100 km/s (typical of DM speeds in galaxies today).  However, if the DM is a thermal relic and annihilation is p-wave dominated then in order to get the correct relic abundance $\sigmav \sim 10^{-31} (10^{-32})$ cm$^3$/s which is many orders of magnitude below the cross section probed by the CMB. On the other hand, the annihilation rate in the WD will be $\sigmav \sim 10^{-30} (10^{-31})$ cm$^3$/s. These rates are probed by WD cooling even when the DM fraction in a GC like NGC 6397 is many orders of magnitude below our constraints and may very well be probed by WDs in a dwarf galaxy such as Segue I if the WD cooling limit could be observed in such objects.

\section{Discussion and Conclusions}
\label{section:conclusion}

Stars accumulate weakly-interacting DM in their cores as they orbit within their host halos. If the dark matter particles annihilate with a cross-section within many orders of magnitude of the canonical thermal relic cross-section of $\langle \sigma_{\mathrm{a}} v\rangle \sim 3 \times 10^{-26}\, \cm^3/\mathrm{s}$, then the dark matter will contribute energy to the stellar core. In most reasonable cases, this energy contribution is a negligible portion of the energy budget of the star. However, in the case of a WD, the luminosity from annihilations can be significant enough to prevent the star from cooling beyond a minimum temperature that can be probed by contemporary or future astronomical observations. Because the annihilation rate is proportional to the local DM density, we can use the coolest WD in a GC to put an upper limit on the DM density in the environment. We have explored constraints on DM required by observations of the coolest WDs in the GC NGC 6397.

This WD cooling argument can be used to constrain a combination of GC structure (parameterized as the fraction of the GC mass in DM, $\fdm$) and dark matter particle scattering cross-section. In particular, we have seen that if the events observed in CoGeNT, DAMA, CRESST, and CDMS-Si could be interpreted broadly as WIMP-like DM, then observations of the WD cooling sequence in NGC 6397 limit $\fdm \lesssim 10^{-3}$ for the best-fit values of $\Mx\ \mathrm{and}\ \sigmaxp$. This would be a significant improvement over existing, kinematic constraints on the DM content of GCs ($\fdm \sim .5$). Additionally, DM WIMPs with parameters consistent with the exclusion limits from LUX and SuperCDMS could still potentially place a powerful constraint on $\fdm$.

This type of WIMP astronomy may have utility far beyond what we point out here. Note that the capture rate of WIMPs in a star is directly proportional to the local density of DM and inversely proportional to the velocity dispersion of the DM distribution. Therefore, the most interesting environments for applying our method have the highest values of the ratio of density to velocity dispersion. Ref.~\cite{Hooper} has already pointed out that this ratio is almost certainly large in Dwarf Galaxies, and so dwarf galaxies can be attractive environments in which to apply such a constraint. Dwarf galaxies have the disadvantage that they are extremely distant from the Solar System, and therefore it is extremely difficult to place a firm lower limit on the luminosities of WDs in these objects. GCs have the advantages that there are a number of relatively dense, nearby clusters, and that GCs are often observed to great depths to further a variety of scientific goals including the study of low-mass stellar evolution, stellar remnants, and cluster structure. 

We define the ratio of the density to the velocity dispersion as  
%
\begin{equation}
\alpha = \frac{c}{\bar{v}},
\end{equation}
where $c$ is the central concentration of the King model for the globular cluster (which can be thought of as a proxy for the central density). In order to constrain $\fdm$ in a given environment, our method requires the observation of a cool WD. In particular, the observations must be sufficiently deep that the coolest, dimmest WDs can be identified with the end of the WD cooling sequence and that no slightly cooler WDs were missed in the observations simply because they are dimmer. Therefore, we should consider nearby GCs such that one could observe the full WD cooling sequence in the cluster. 
%
\begin{figure}[htp]
\centering
\includegraphics[width=9cm, height=9cm]{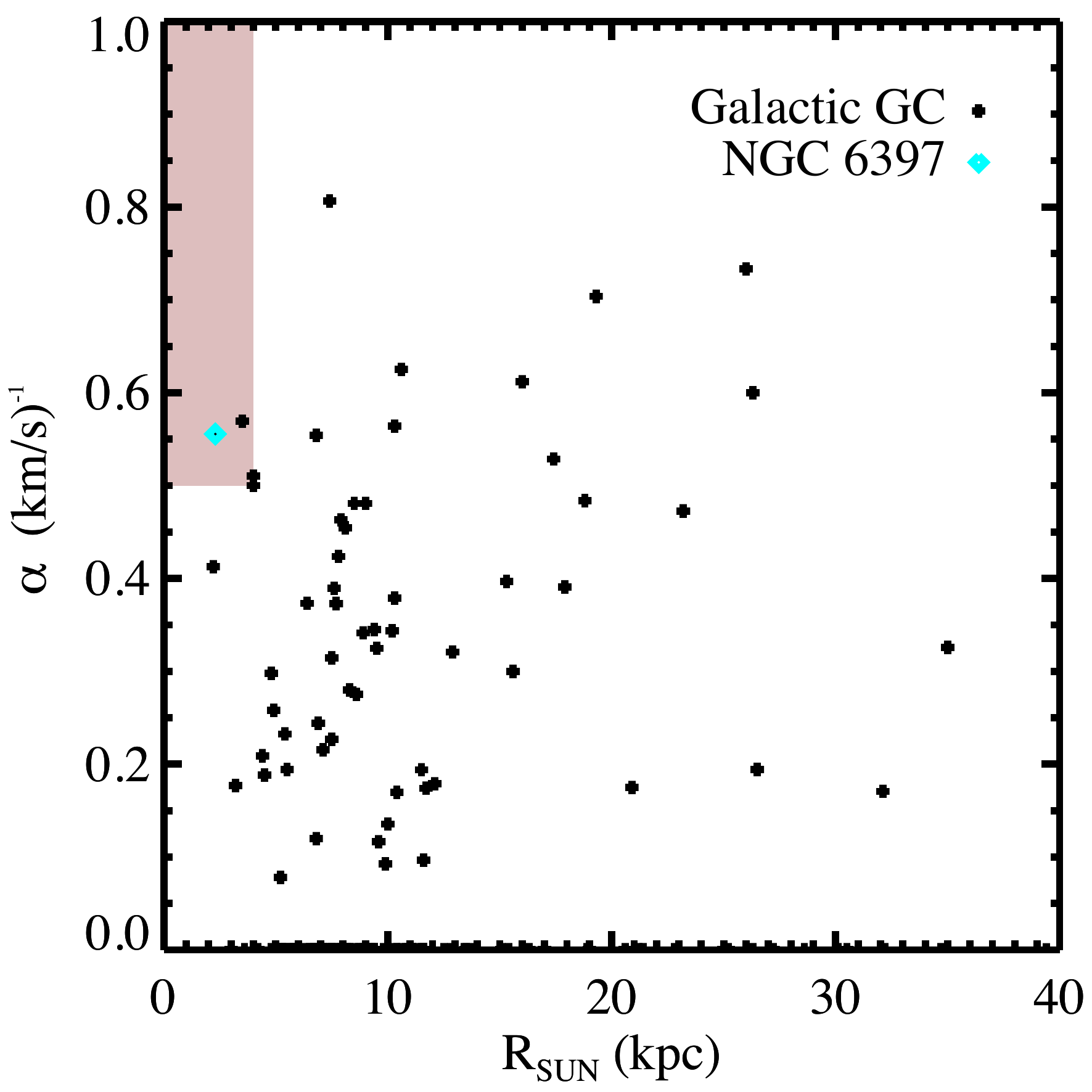}
\caption{The galactic globular cluster distribution.  $\alpha$ is the ratio of the King model central concentration to the velocity dispersion of the cluster and $\Rsun$ is the distance from the Sun to the cluster.  The black dots represent galactic GCs while the cyan diamond is for NGC 6397.  The shaded region represents a hypothetical selection criterion for GCs.  GCs in this region should be interesting candidates for application of our method if high quality observations of the WD cooling sequence are obtained.}
\label{fig:globulars}
\end{figure}

Fig.~\ref{fig:globulars} shows the distribution of Galactic GCs. If we consider GCs with $\alpha$ comparable to or greater than that of NGC 6397 [say $\alpha \gtrsim 0.5\, (\mathrm{km/s})^{-1}$] located not too much further than NGC 6397 (say within 4 kpc of the Sun), then we have 3 candidates for further observation:  NGC 6366 ($\alpha = 0.57, \Rsun = 3.5\, \kpc$), NGC 6752 ($\alpha = 0.51, \Rsun = 4.0\, \kpc$) and NGC 6838 ($\alpha = 0.50, \Rsun = 4.0\, \kpc$) \cite{Harris}.  Of these, NGC 6366 and NGC 6838 are quite different than NGC 6397 in that they are neither metal-poor nor post core-collapse clusters, making them all the more intriguing as targets \cite{Harris}.  NGC 6366 is the most different from NGC 6397 in that it appears to have been heavily tidally stripped \cite{Paust}.  Thus, it is not a particularly dense cluster at all (c = 0.74) but is included in our cut because of its correspondingly low velocity dispersion ($\bar{v} = 1.3 \pm 0.5$ km/s) \cite{Harris}.  Meanwhile NGC 6752 is the most similar to NGC 6397, though it is not as metal poor \cite{Harris}.  All 3 of these GCs have been observed by HST \cite{Paust,Huang,Thomson}; however, it appears that only NGC 6752 has been imaged deeply enough to potentially identify the truncation of the WD cooling sequence (see Fig. 2 in Ref~\cite{Thomson} for the color-magnitude diagram).

In addition to GCs, we could potentially apply our method to the dwarf satellite galaxies of the MW.  These satellites are known to be DM dominated objects, with Segue~I having the highest central DM density at $\rho_\chi \sim 100\, \gev/\cm^3$ \cite{Simon}.  The hypothetical observation of a truncated WD cooling sequence in a dwarf galaxy such as Segue~I could be uniquely constraining, as pointed out in Ref.~\cite{Hooper}. In Fig.~\ref{fig:mvsigma2}, we reiterate this point and extend previous results to low-mass DM candidates in order to show that such observations have the unique ability to probe low-mass DM, a point which was missed in Ref.~\cite{Hooper}.  The cyan dot-dashed curve in Fig.~\ref{fig:mvsigma2} shows what the constraint would be if a WD was observed in Segue~I with a temperature of 3500-3700 K.  Because the DM density in Segue~I is well constrained, observation of a cool WD within the half-light radius of this satellite could potentially rule out a broad range of parameter space extending to low WIMP masses and annihilation rates far below the canonical thermal value.

There is a long history of interplay between particle physics and astrophysics in understanding the fundamental constituents of the universe and the laws that govern their interactions. In this paper, we explore the possibilities for a unique interplay between the physics of the dark matter and the evolution of star clusters. In particular, we show that possible near-future measurements of the properties of the dark matter may immediately result in profoundly more stringent constraints on models of globular cluster formation. Conversely, if there were ever a strong astrophysical reason to suspect that GCs formed in significant DM halos, then the evolution of WDs in GCs places strong constraints on dark matter. These constraints are relevant to dark matter masses and annihilation cross section far below those that can be probed using any other techniques. Indeed, we have shown that it may be possible in the future to extend the results of Ref.~\cite{Hooper}, such that observations of WDs in nearby dwarf galaxies may constrain dark matter at masses up to three orders of magnitude below what may be attainable with contemporary direct or indirect dark matter searches. This may point the way toward a new frontier for exploration at the interface between particle physics and astrophysics.

\vspace*{12pt}
\begin{acknowledgments}
TJH would like to thank Brad Hansen and Jason Kalirai for their help in locating and interpreting the observational data for NGC 6397. We are also grateful to Dan Hooper and Savvas Koushiappas for helpful email exchanges regarding this work.  The work of TJH and ARZ was supported, in part, by the Pittsburgh Particle physics, Astrophysics, and Cosmology Center (Pitt-PACC) at the University of Pittsburgh and U.S. National Science Foundation Grant NSF PHY 0968888.  A.N. acknowledges funding from NASA ATP grant NNX14AB57G. A.N. thanks Pitt-PACC and the Department of Physics and Astronomy at the University of Pittsburgh for partial financial support.
\end{acknowledgments}

\bibliography{dmwdgc}

\end{document}